\documentstyle[prl,aps,epsfig,prb,floats]{revtex}
\begin{document}
\draft 
\title{A numerical exact solution of the Bose-Hubbard model}
\author{N.~Elstner and  H.~Monien} 
\address {Physikalisches Institut, Universit\"at Bonn, 
          Nu\ss allee 12, D-53115 Bonn, Germany}

\twocolumn[ \date{\today} \maketitle \widetext
\begin{abstract}
\begin {center}
  \parbox{14cm}{In this paper we report results from a systematic 
   strong-coupling expansion of the Bose-Hubbard model in one and
   two spatial dimensions. We obtain numerically exact results for 
   the structure factor and the spectrum of single particle and 
   single hole excitations in the Mott insulator. This enables the
   determination of the zero-temperature phase diagram and the 
   location of the critical endpoints of the Mott lobes. In one 
   dimension we confirm the occurrence of reentrance behavior from
   the compressible to the insulating phase in a region close to
   the critical point.
}
\end{center}
\end{abstract}
\pacs{\hspace{1.9cm} PACS numbers: 05.30.Jp, 05.70.Jk, 67.40.Db} ]
\narrowtext

\section{Introduction}

Quantum phase transitions in strongly correlated systems have attracted
a lot of interest in recent years. Here we focus on systems of interacting 
bosons. Physical realisations include Josephson junction arrays, granular 
and short-correlation-length superconductors, flux-lattices in type-II 
superconductors and possibly in the future ultracold atoms in a periodic
potential. 

The minimal model containing the key feature of competition between
kinetic and potential energy is the Bose-Hubbard hamiltonan. It is
given by
\begin{equation}
H = -t\sum_{<i,j>} \left(b_i^\dagger b_j + b_j^\dagger b_i \right) 
    + U \frac{1}{2} \sum_i \hat n_i(\hat n_i - 1) - \mu \sum_i \hat n_i
\label{bh-model}
\end{equation}
where the $b_i^\dagger$ and $b_i$ are bosonic creation and
annihilation operators, $\hat n_i = b_i^\dagger b_i$ is the number of
particles on site $i$, $t$ the hopping matrix element, $U>0$ the
on-site repulsion and $\mu$ the chemical potential. 

This model has been studied previously by Quantum Monte Carlo simulations 
\cite{SBZ91,BS92,NSFG94,KT91,KTC91,OW94,BSZK95} in one and two spatial
 dimensions. Recently the one-dimensional case was also investigated 
using the density-matrix renormalisation group (DMRG). This study found
indications for reentrence from the superfluid to the Mott-insulator for 
certain parameter values. 

The zero-temperature phase diagram can be understood by starting from 
the strong-coupling or ``atomic'' limit where the the kinetic energy
term vanishes ($t=0$). In this limit the ground state is an insulator
with a fixed number $n_0$ of particles per site and the wavefunction
is given by
\begin{equation}
\left| n_0 \right>_{\rm Mott}^{(0)} = \prod_{i=1}^N \frac{1}{\sqrt{n_0!}}
                                \left(b^\dagger_i\right)^{n_0} \left|0\right> 
\end{equation}
with energy
\begin{equation}
	E_{\rm Mott}^{(0)}/N = U \frac{1}{2} n_0(n_0 - 1) - \mu n_0 \;\; .
\end{equation}
Single charge excitations in the atomic limit are created by adding or
removing a particle onto or from a particular site $i$:
\begin{eqnarray}
\left| n_0;i \right>_{\rm part}^{(0)} &=& 
                                     \frac{1}{\sqrt{(n_0+1)}} \, b^\dagger_i 
                        \left| n_0 \right>_{\rm Mott}^{(0)} \label{eq:sp-state} \\
\left| n_0;i \right>_{\rm hole}^{(0)} &=& \frac{1}{\sqrt{n_0}} \, b_i
                                    \left| n_0 \right>_{\rm Mott}^{(0)} 
                                    \label{eq:sh-state}
\end{eqnarray}
The energy of these states relative to the ground state is given by
\begin{eqnarray}
	E_{\rm particle}^{(0)} &=& U n_0 - \mu \label{eq:sp-energy} \\
	E_{\rm hole}^{(0)} &=& -U ( n_0 -1 ) + \mu \label{eq:sh-energy}
\end{eqnarray}
for particle and hole excitations respectively. For certain critical 
values of the chemical potential $\mu_c^{(0)} = U n_0$ this energy 
vanishes and the system becomes compressible. 
If the hopping matrix element $t$ is finite the range of the chemical 
potential for which the system is incompressible decreases. At some
critical value $t_c$ of the hopping matrix element the Mott insulator
will completely disappear. The resulting phase diagram shows Mott 
insulating regions, usually referred to as Mott lobes, surrounded by 
a compressible phase.

This phase is characterized by the following real space correlation function:
\begin{equation}
S_j = \left< b_j^\dagger b_0 + b_j b_0^\dagger \right> 
\end{equation}
In spatial dimension $D>1$ the correlations are long ranged and the system
is a superfluid, while in one dimension it is a Luttinger Liquid with 
$S_j$ decaying algebraically for large distances. 
In this article we will investigate the Fourier transform 
\begin{equation}
S({\bf q}) = \sum_j e^{i {\bf q} \cdot {\bf r}_j} \, S_j
\end{equation}
and it's second moment
\begin{equation}
m_2({\bf q}) = -\frac{\partial^2}{\partial{\bf q}^2} S({\bf q})
\end{equation}
Assuming a lorentzian shape of $S({\bf q})$ near the ordering wave vector
${\bf q}=0$ the correlation length, $\xi$, is given by
\begin{equation}
\xi^2 = \frac{m_2({\bf q}=0)}{S({\bf q}=0)} \;\;\;\;.
\end{equation}

We used strong coupling expansions in $t/U$ to calculated the ground state 
energy, $E_0$, the spectrum of single charge excitations 
$E_{particle/hole}({\bf q})$, the correlation function $S({\bf q})$ and the
correlation length $\xi$. The resulting series were analysis using 
Pade extrapolation techniques.

In the next section we give a technical outline of how to calculated 
the ground state energy, $E_0$, the spectrum of single charge excitations 
$E_{particle/hole}({\bf q})$, the correlation function $S({\bf q})$ and the
correlation length $\xi$ by strong coupling expansions in $t/U$. We use 
methods that had been developed over the last decade but so far were
mainly applied to spin systems.

Section III focusses on the ground state and the excitated states. We
present the dispersion of the single particle excitations and the dependence
of the particle-hole gap on the hopping amplitude $t$. It is found that the
series converges rather fast provided $t$ is smaller then the critical value
$t_c$. In this regime, the Mott insulator, we obtain the numerical exact 
single charge excitation spectrum.

In the following section the phase diagram in two dimensions is discussed. 
Again due to the rapid convergence of the series it is possible to 
determine numerically exact the boundary of the Mott lobes. Only in a small 
region close to the critical point $t_c$ does the convergence break down 
and series extrapolation techniques have to be applied. These also work 
extremely well and it is thus possible to determine the critical point
with high accuracy. We also determine the critical exponents and
find that they agree with the field theoretic predictions that this
$D$-dimensional quantum system is in the universality class of the 
the $D+1$-dimensional classical xy-model.

Section V deals with the special case of a one-dimensional system. 
As already mentioned the compressible phase exhibits only quasi 
longrange order. Also the phase transitions at the top of the Mott lobes
changes qualitatively and becomes of the Bereshinski-Kosterlitz-Thouless
(BKT) type. Our analysis of the phase diagram finds a reentrance behaviour,
i.e. for a certain range of values of the chemical potential the system has 
not just one transition from the Mott state to the compressible phase 
but upon further increase of the hopping  strength $t$ becomes insulating
again until a second transition to the liquid occurs. This scenario agrees 
with recent DMRG calculations. We give a simple intuitive explanation 
for this surprising observation.

Conclusion

\section{Series Expansion Technique for Excited States}

Strong coupling perturbation theory can be formulated as a linked 
cluster expansion, see e.g. \cite{GSH90}. For the ground state energy
this is basically nondegenerate Rayleigh-Schr\"odinger perturbaton
theory performed for every cluster contributing up to a certain order
in the expansion. In order to obtain correlation function one has
to formaly add  source terms,
\begin{equation}
H_{\rm source} = \sum_{i,j} J_{i,j} \, 
\left( b_i^\dagger b_j + b_i b_j^\dagger \right) \, \;\;\;\;, 
\end{equation}
to the hamiltonian and expand the energy to linear order in
the ${J_{i,j}}$. Differentiating with respect to the source
fields will give the strong coupling expansion for the 
correlation functions. This procedure is necessary, because it is not
possible to evaluate the wave function directly by a linked 
cluster expansion. This is due to the fact that the linked cluster
theorem applies only to physical observables that are additive
when the system seperates into disconnected part.

A systematic strong coupling expansion of the energy of the charge
excitations complicated due to the high degeneracy. The problem how to
write down a linked cluster expansion for degenerate states was solved
only recently by Gelfand\cite{Gelfand:96}. The idea is to construct
perturbatively an effective Hamiltonian $H^{\rm eff}_{i,j}$ in the
subspace of the degenerate states $\left|n_0;i\right>_{\rm
  part/hole}^{(0)}$ by a similarity transformation
\begin{eqnarray}
H^{\rm eff}_{i,j}(t) &=& S_{i,\nu}(t) \, H_{\nu,\lambda} \, S_{\lambda,j}(t) \\
{\rm with} \phantom{W} S_{i,\nu}(t) &=& S_{\nu,i}^{-1}(t) \nonumber
\end{eqnarray}
where Greek indices run over states in the full Hilbert space while
Latin indices are restricted to the degenerate manifold of single
particle and single hole states (\ref{eq:sp-state}) and (\ref{eq:sh-state})
respectively. Then the linked cluster theorem applies to $H^{\rm
  eff}_{i,j}(t) - E_{\rm Mott}(t)$. In the case of a homogeneous system
$H^{\rm eff}_{i,j}$ depends only on the difference of indices $i-j$
and is easily diagonalised by a Fourier transform. This way one can
determine the full dispersion $E({\bf q}; t,\mu)$ of the charge
excitations. In many ways the linked cluster expansion is similar to a
exact diagonalization study of small systems - however in the linked
cluster expansion it is possible to remove all finite size effects in
each order and one obtains the full dispersion in the thermodynamic
limit. 

We close this section with some technical remarks. Expansions were performed 
up to 
13$^{th}$ order for the triangular lattice with 12,253 clusters contributing.
The Hilbert-space of the largest cluster had 37,442,160 states in the
the single particle sector with filling $n_0 = 1$. Up to 13$^{th}$ order 
in perturbation theory just a fraction of these states (788,238)  actually 
contribute to the expansion. It was only after we restricted the calculation 
to this subset of relevant states that it was possible to perform this work.
Any significant improvement is unlikely because the Hilbert-space and 
the number of clusters both increase exponentially with the order of the 
expansion. 

\begin{figure}[t]
  \begin{center} \epsfig{file=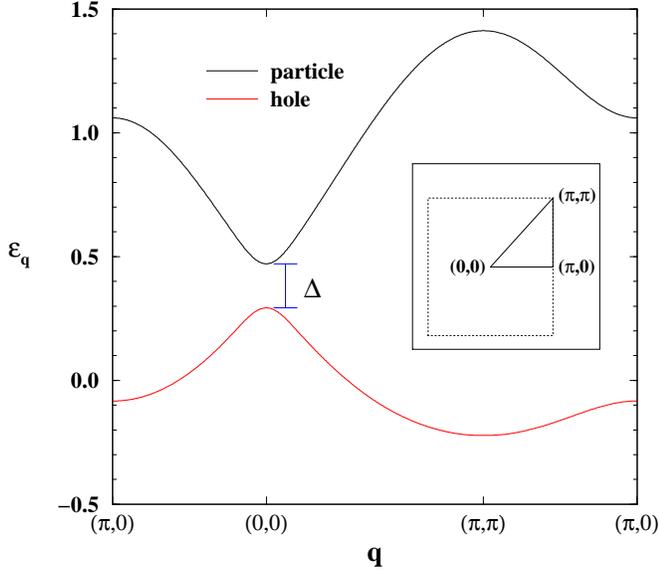,width=246pt} 
  \caption[The particle and hole excitation gap] 
          {Dispersion of the single-particle and single-hole excitations
           of the square lattice Bose-Hubbard model at $t/U=0.055$}
  \label{Dispersion-sqr} \end{center}
\end{figure}

\begin{figure}[t]
  \begin{center} \epsfig{file=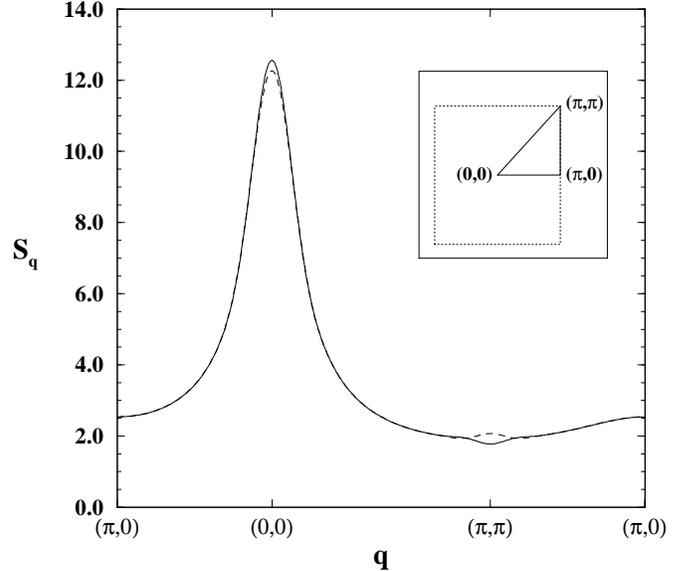,width=246pt} 
  \caption[The correlation function] 
          {Correlation function $S(k)$ of the square lattice
           Bose-Hubbard model at $t/U=0.055$}
  \label{Dispersion-corr-sqr} \end{center}
\end{figure}

\section{Ground State and Single Charge Excitations}

We will first discuss the two dimensional case. We investigated both
the square and triangular lattice and calculated the series for
occupation numbers $n_0=1$ and $n_0=2$ up to 13$^{\rm th}$ and
10$^{\rm th}$ order respectively. 

The spectrum $E({\bf q}; t,\mu)$ takes on the form
\begin{eqnarray}
  E_{\rm part}({\bf q};t,\mu) &=& \epsilon_{\rm part}({\bf q};t) - \mu \\
  E_{\rm hole}({\bf q};t,\mu) &=& -\epsilon_{\rm hole}({\bf q};t) + \mu 
\end{eqnarray}
in complete analogy to eqns. (\ref{eq:sp-energy}) and 
(\ref{eq:sh-energy}). 
For positive values of the hopping matrix element $t$ the smallest
(largest) eigenvalue in the particle (hole) sector is always located
at wavevector ${\bf q} = 0$. 
The upper and lower phase boundaries of the Mott phase 
are thus given by 
$\mu_{\rm upper}(t) = \epsilon_{\rm part}({\bf q}=0;t)$ and 
$\mu_{\rm lower}(t) = \epsilon_{\rm hole}({\bf q}=0;t)$, respectively.

As a consequence the single charge gap 
$\Delta(t) = \epsilon_{\rm part}({\bf q}=0;t) 
- \epsilon_{\rm hole}({\bf q}=0;t)$, determines also the width 
$\mu_{\rm upper}(t) - \mu_{\rm lower}(t)$ of the insulating region. 
With increasing hopping $t$ the distance between the upper and lower
boundary decreases until finally at some critical value, $t_c$, the
energy to remove a particle and the energy to add a particle become
degenerate and the Mott insulator vanishes altogether.

\begin{figure}[t]
  \begin{center} \epsfig{file=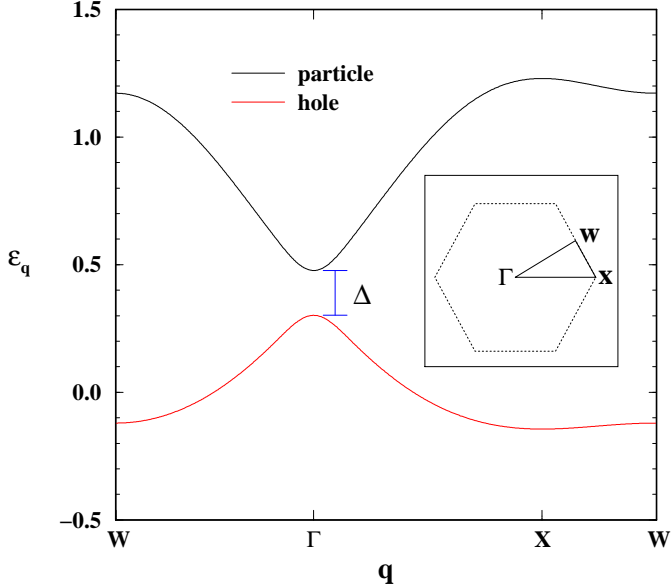,width=246pt} 
  \caption[The particle and hole excitation gap] 
          {Dispersion of the single-particle and single-hole excitations
           of the triangular lattice Bose-Hubbard model at $t/U=0.035$}
  \label{Dispersion-tri} \end{center}
\end{figure}

The dispersion of the particle and hole excitations for $n_0=1$ on the
square lattice and triangular lattice is shown in Figs.
\ref{Dispersion-sqr} and \ref{Dispersion-tri}. The different shape of 
the two curves reflects the particle-hole asymmetrie of the model 
hamiltonian (\ref{bh-model}). 
The series were found to converge very fast. Both Figures 
were obtained by summation of the 13$^{\rm th}$ order series. 
They turned out to be almost indistinguishable from the 
result of the 10-term series even for $t/U=0.055$ (square lattice) and 
 $t/U=0.035$ (triangular lattice) which is not far from the critical 
endpoints $t_c$ of the first Mott lobe for these two lattices. 
The particle and hole excitations both have a pronounced extremum at 
wavevector ${\bf q}=0$ and are separated by a gap $\Delta$.  For values 
of the chemical potential $\mu$ in this range all single charge excitations 
are gapped and the system is insulating. Coefficients of the series expansion 
of $\Delta$ are given in Tab.(1). 

\begin{figure}[t]
  \begin{center} \epsfig{file=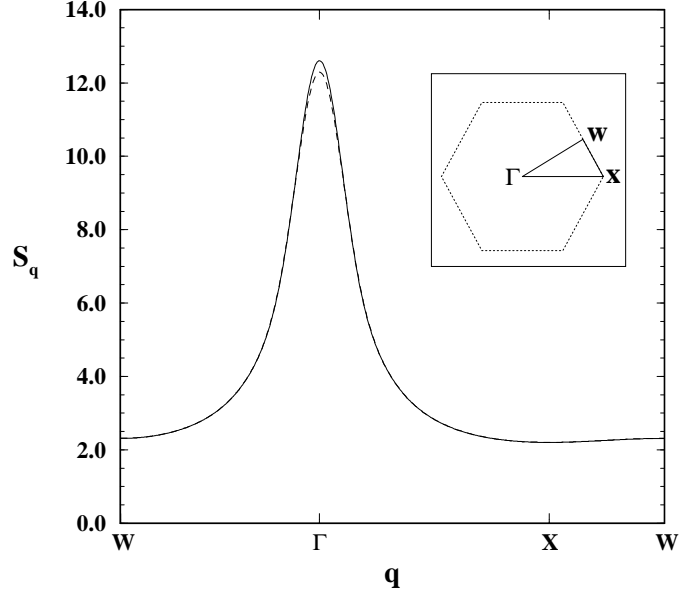,width=246pt} 
  \caption[The correlation function] 
          {Correlation function $S(k)$ of the triangular lattice
           Bose-Hubbard model at $t/U=0.035$}
  \label{Dispersion-corr-tri} \end{center}
\end{figure}

Figs.\ref{Dispersion-corr-sqr} and \ref{Dispersion-corr-tri} show the 
correlation function $S({\bf q})$ for the same sets of parameters as for the
dispersion. Although the system is still in the
Mott phase there is already a prounced peak at wavevector ${\bf q} = 0$.

\section{2D Phase Diagram}

\begin{figure}[t]
  \begin{center} 
        \epsfig{file=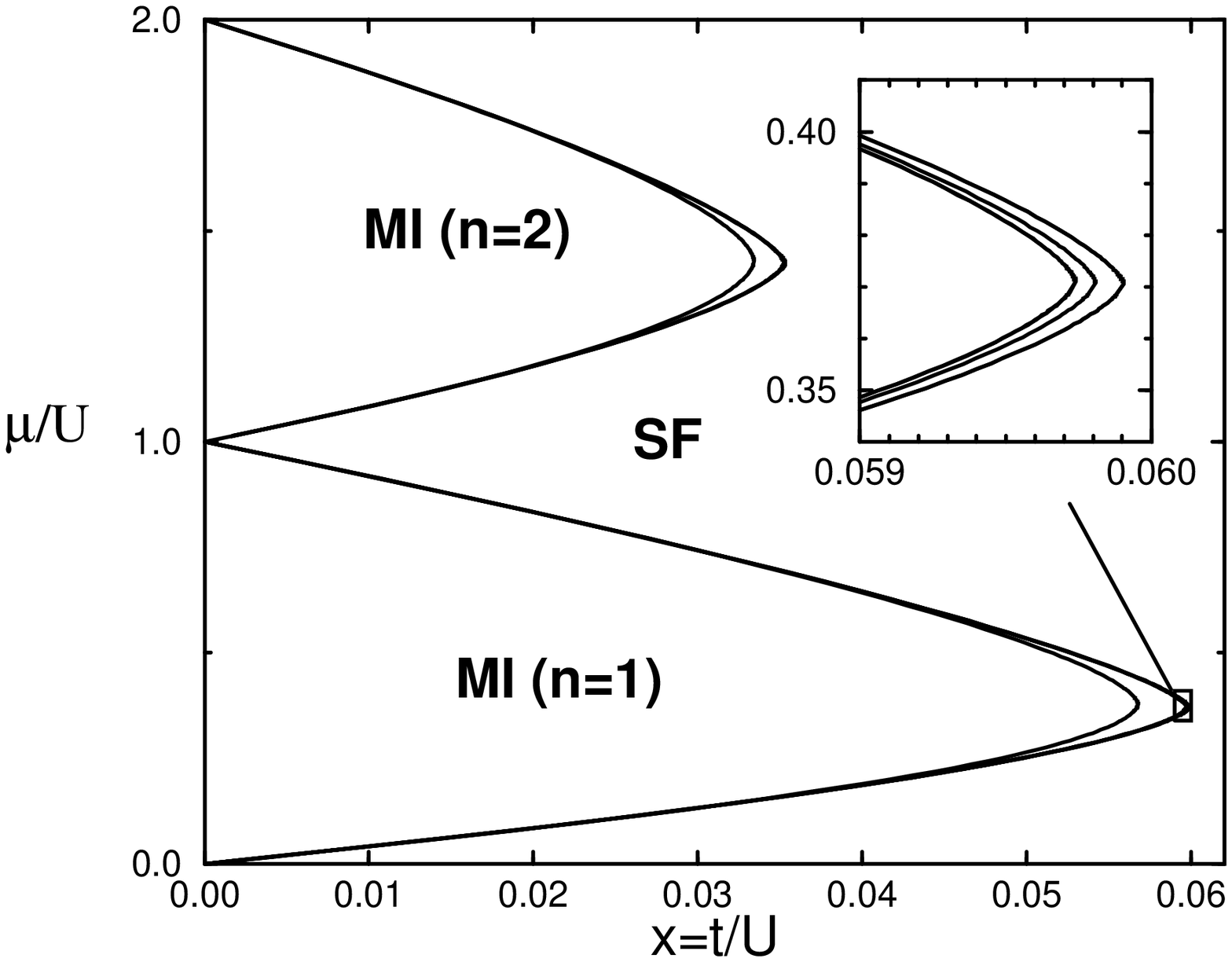,width=246pt} 
        \caption[Phase Diagram] 
                {Square lattice: Phase Diagram obtained by Pade
                 analysis.}
        \label{phase-diagram:sqr} 
   \end{center}
\end{figure}

The phase diagram can be obtained from the series for the single 
charge exciations. Outside the critical region the values of the 
chemical potential for which the system becomes compressible can 
be directly calculated from the series. Only near the critical point 
is it necessary to use extrapolatation methods. We applied a Pade 
analysis and obtained the phase diagrams shown in 
Figs.\ref{phase-diagram:sqr} and \ref{phase-diagram:sqr} for the 
square and triangular lattices.
With the exception of the lowest approximand all others turn out 
to be almost indistinguishable again indicating rapid convergence. 

\begin{figure}[t]
  \begin{center} 
        \epsfig{file=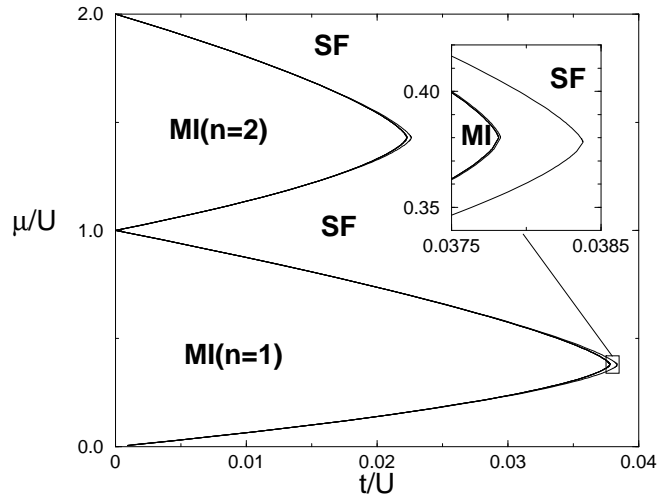,width=246pt} 
        \caption[Phase Diagram] 
                {Triangular lattice: Phase Diagram obtained by Pade
                 analysis.}
        \label{phase-diagram:tri} 
   \end{center}
\end{figure}

\begin{figure}[t]
  \begin{center} 
        \epsfig{file=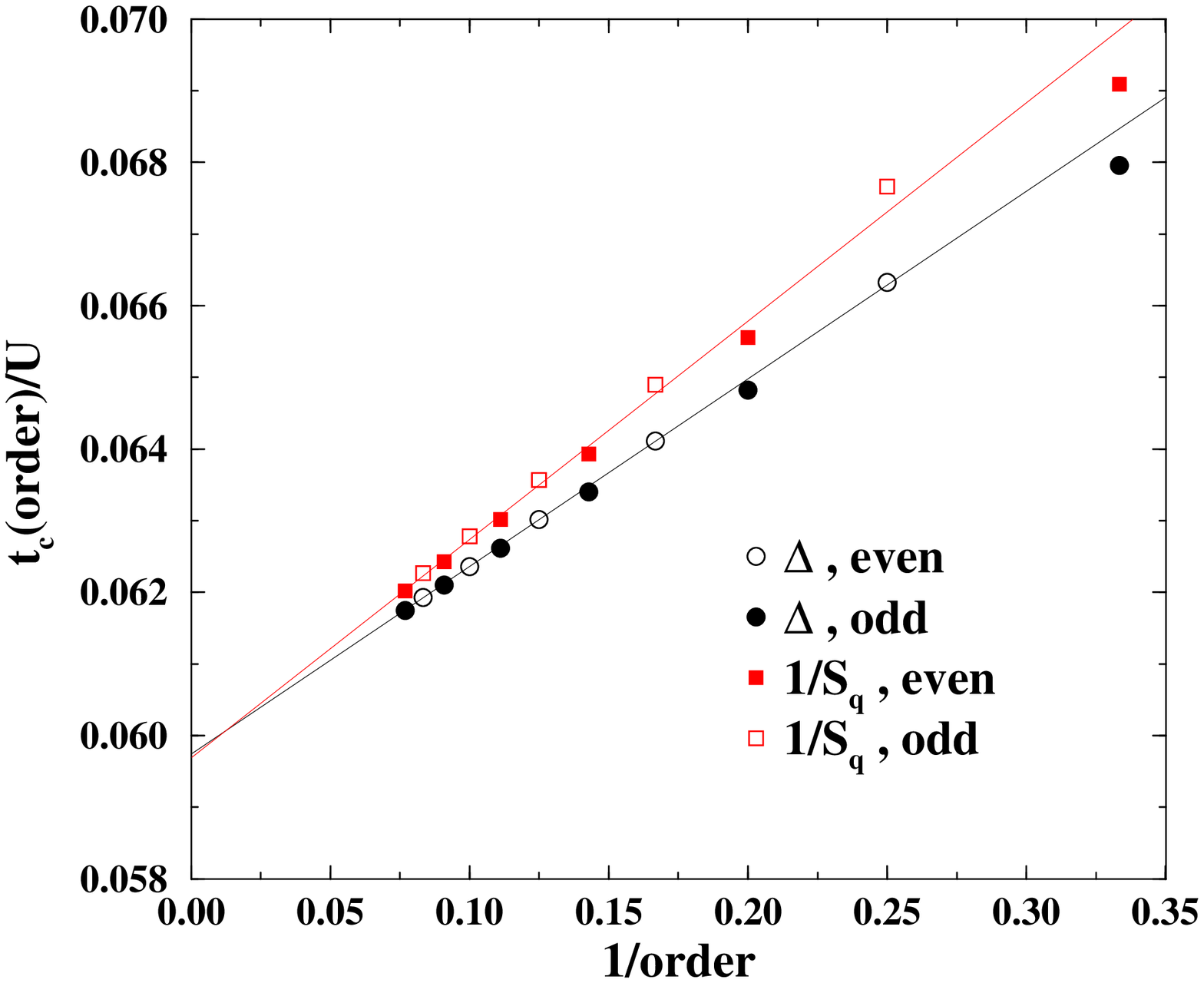,width=246pt} 
        \caption[The critical point] 
                {Square lattice: $1/k$ extrapolation of the critical 
                 point $t_c$. The circles and squares are estimates from 
                 series for the gap and correlation function respectively.
                 Note the excellent agreement between the $t_c$ estimates 
                 of these two independendant quantities.
}
        \label{crit-point0:sqr} 
   \end{center}
\end{figure}

\begin{figure}[t]
  \begin{center} 
        \epsfig{file=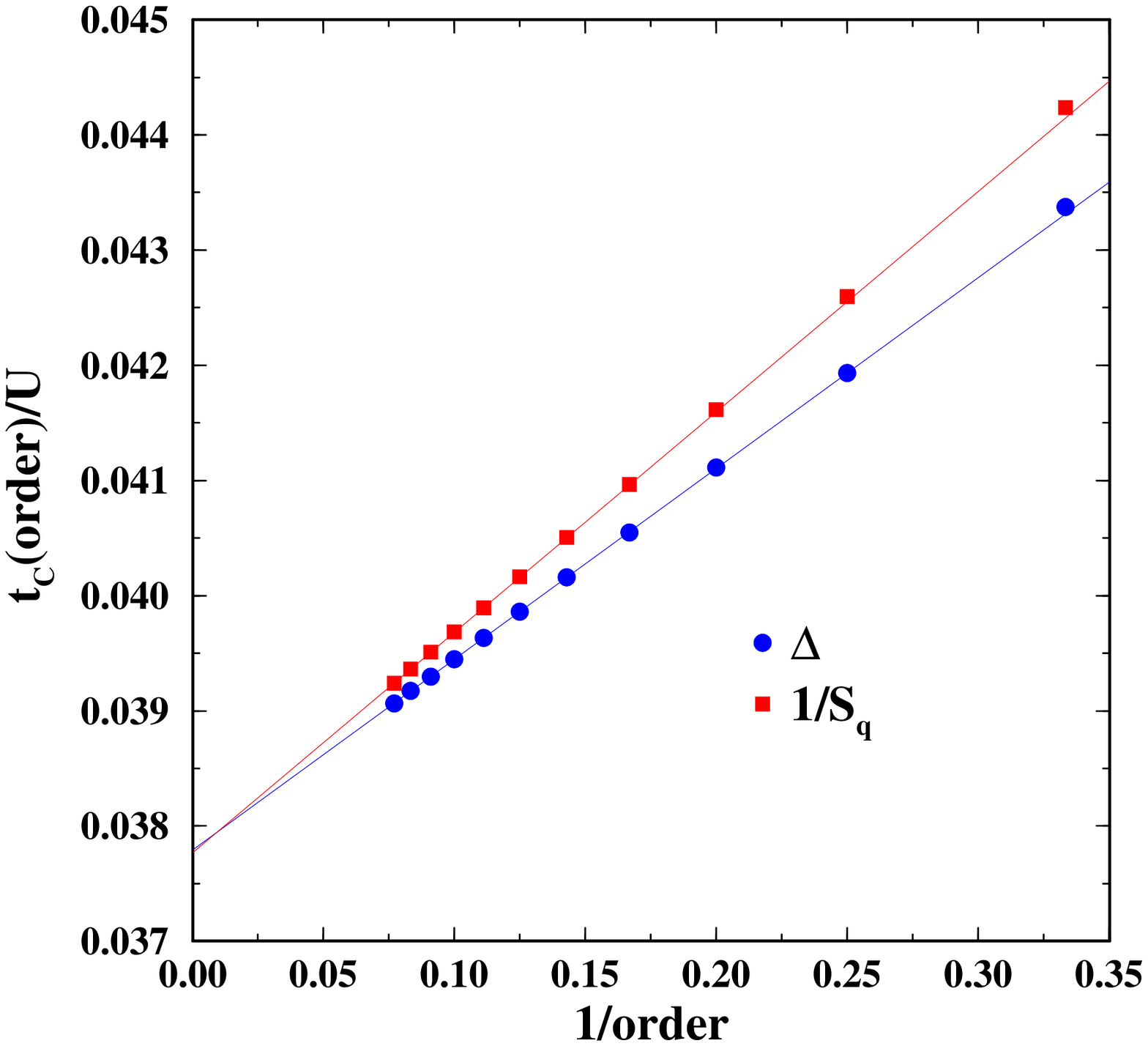,width=246pt} 
        \caption[The critical point] 
                {Triangular lattice: $1/k$ extrapolation of the critical 
                 point $t_c$. The circles and squares are estimates from 
                 series for the gap and correlation function respectively.
                 Note the excellent agreement between the $t_c$ estimates 
                 of these two independendant quantities.}
        \label{crit-point0:tri} 
   \end{center}
\end{figure}

\begin{figure}[t]
  \begin{center} 
        \epsfig{file=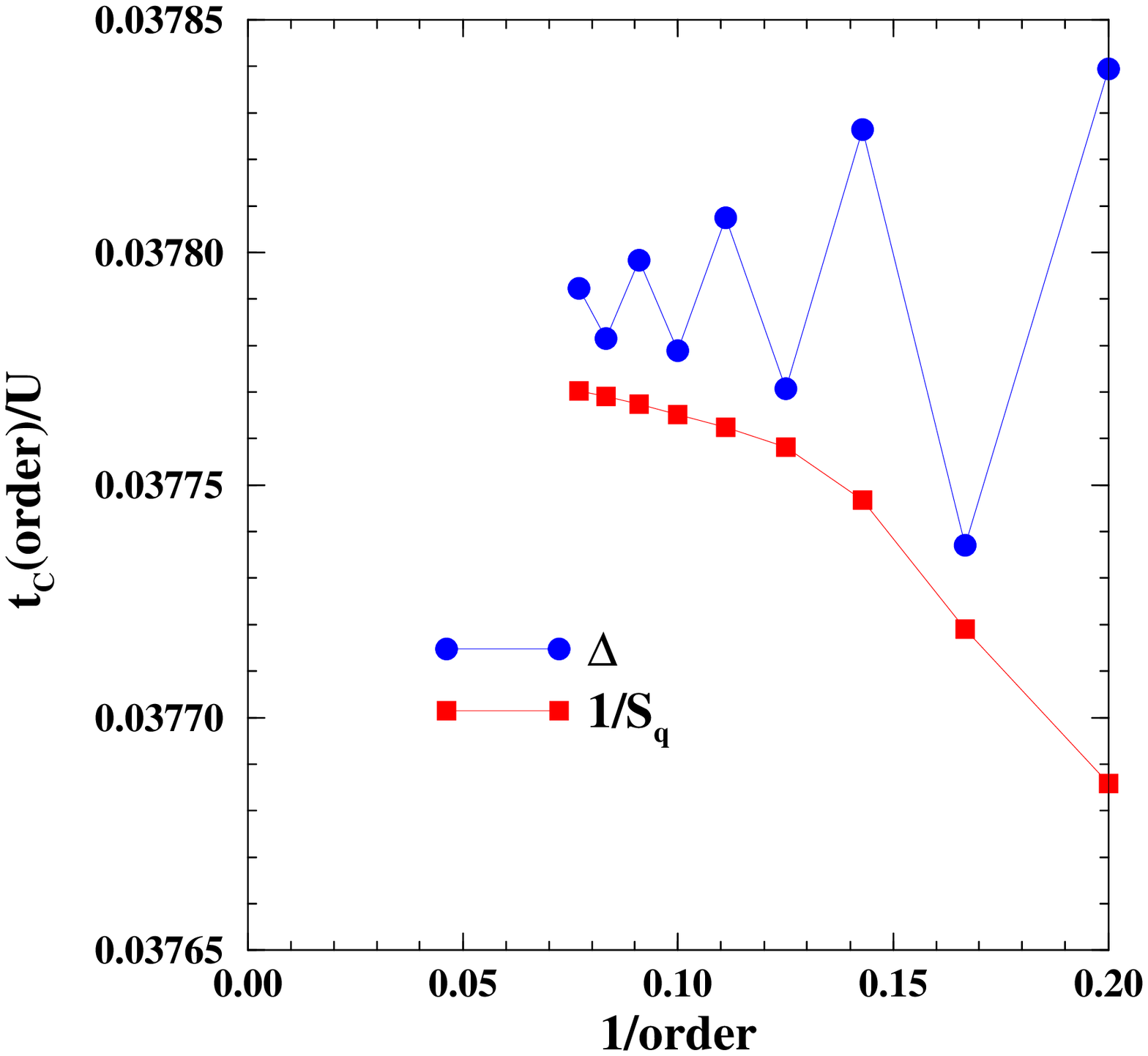,width=246pt} 
        \caption[The critical point] 
                {Triangular lattice: $1/n$ extrapolation of the critical 
                 point sequence $t_c^{(1)}$. Note the scale of the 
                 vertical axis!}
        \label{crit-point1:tri} 
   \end{center}
\end{figure}

The bare series shows a remarkable fast convergence for not too large values 
of the expansion parameter. In each order $k$ of the expansion $\Delta$ 
vanishes at some effective critical value $t_c{(k)}$ with a corresponding 
effective $\mu_c{(k)}$. Plotting $t_c{(k)}$ and $\mu_c{(k)}$ vs. $1/k$ 
one finds again a rapid convergence as shown in Figs.(\ref{crit-point0:sqr}) 
and (\ref{crit-point0:tri}). 
Extrapolation to $n\rightarrow\infty$ allows to determine accurately the 
critical point: 
\begin{eqnarray}
t_c = 0.05974\phantom{0} \pm 0.00004\phantom{0} 
&\hskip 24pt& (\rm square\; lattice) \\
t_c = 0.037785 \pm 0.000005 &\hskip 24pt& (\rm triangular\; lattice) 
\label{tc:1/k}
\end{eqnarray}
The accuracy for the triangular lattice is almost an order of magnitude 
better then for the square lattice. This is because the series for the 
triangular lattice has a monotonic convergence while the square lattice 
shows an oscillatory behaviour between even and odd terms as can be seen
when comparing Figs.(\ref{crit-point0:sqr}) and (\ref{crit-point0:tri}).
The convergence of the data is significantly accelerated when considering
the difference between neighbouring data points. Fitting each pair of 
points $t_c(k)$ and $t_c(k-1)$ to a linear function 
$t_c^{(1)}(k) + a(k)/k$ removes the leading $1/k$ dependance and gives
an even faster converging new sequence of points $t_c^{(1)}(k)$. For the 
triangular lattice the result is shown in Fig.(\ref{crit-point1:tri}). 
Whithout further extrapolation this confirms the above value
of the critical point.

\begin{figure}[t]
  \begin{center} 
        \epsfig{file=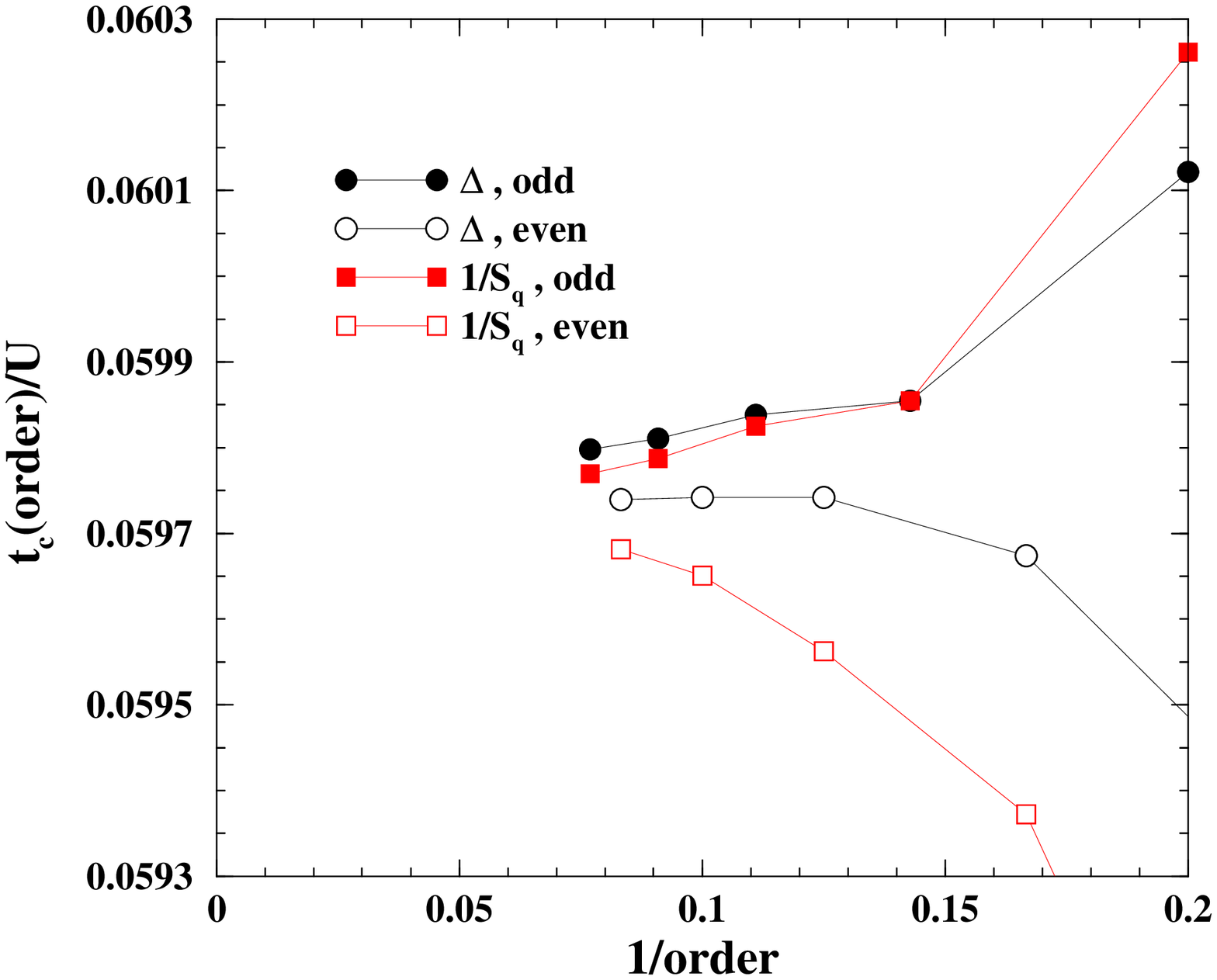,width=246pt} 
        \caption[The critical point] 
                {Square lattice: $1/n$ extrapolation of the critical 
                 point sequence $t_c^{(1)}$. Note the scale of the 
                 vertical axis!}
        \label{crit-point1:sqr} 
   \end{center}
\end{figure}

Having located $t_c$ we now turn to a more detailed investigation of
the critical behaviour. Scaling theory predicts this phase transition 
to be in the universality class of the three-dimensional 
xy-model\cite{Fisher:1989}. The Mott lobe is thus expected to close 
following a power law
\begin{equation}
      \Delta(t) \propto (t_c-t)^{z\nu} \;\; , \;\; (t_c-t) \ll 1 \;\; .
\end{equation}
where $z$ the dynamical exponent is given by $z=1$. From the 
scaling form of the dynamic susceptibility\cite{Fisher:1989} the
power law divergence of the static structure factor can be derived
\begin{eqnarray}
      S(q=0,t) &\propto &  (t_c-t)^{\gamma_s} \;\; , \;\; (t_c-t) \ll 1 \\
{\rm with} \;\; \gamma_s &= & (1-\eta)\nu
\end{eqnarray}
We used a
Dlog-Pade analysis to check these scaling predictions. For the square 
lattice this gives for the critical point $t_c \approx 0.0598 $ and 
the critical exponents $\nu \approx 0.69$ and $\gamma_s = 0.653$. 
The different Pad\'e approximants differ quite a bit and the $t_c$ value 
clearly does not agree with the result of the $1/n$ extrapolation. 
The values for the critical exponent $\nu$ have to be compared with the 
known value for the 3D xy-model\cite{GZ77}, $ \nu = 0.6693 \pm 0.0010 $ 
and $ \gamma_S  = 0.64.. \pm 0.0022 $ , obtained by Borel summation of field 
theoretical results. Also the difference between the two results is 
quite small of the order of a few per cent this is still closer to the 
3D Heisenberg model with $\nu = 0.7054 \pm 0.0010 $\cite{GZ77} then to 
the xy-result. The convergence for the square lattice is not that good 
as mentioned before. When going to triangular lattice the accuracy of 
the analysis improves significantly. 
The Dlog-Pade approximands find the critical point at $t_c \approx 0.03778 $ 
with exponent $\nu = 0.681 - 0.683 $ and $\gamma_s = 0.651 - 0.652$.
The values for different approximants are presented in Tab. \ref{pade:tri}. 
The $t_c$ value is in excellent agreement with the $1/n$ extrapolation. 
The critical exponent still remains higher then the field theoric value. 
Looking carefully at the values for $t_c$ and $\nu$ as given by different 
approximants there is a tendency for both quantities to decrease slowly
with increasing length of the series. In addition to the standard 
Dlog-Pade analysis we also performed a biased analysis with the critical 
point fixed at $t_c = 0.03778$. This gave $\nu = 0.6805 -  0.6810$. 

The critical exponent depends linearly on the critical point. Fixing $\nu$ 
at its field theoretic value this would require $t_c = 0.03775 - 0.03776$ 
which can clearly be ruled out by the $1/n$ extrapolation. 

The discrepancy is probably due to subleading corrections to the asymptotic 
divergent term. The  Dlog-Pade analysis ignores these contributions
and assumes a pure power law divergence at the critical point. An improved 
analysis to resolve this subtle question however would require much longer 
series which won't be possible, because the numerical problem grows 
exponentially with the order of the series.

\section{1D Phase Diagram}

We now turn to the one-dimensional case. Scaling theory\cite{Fisher:1989}
predicts the critical behavior of the system to be that of a 
Kosterlitz-Thouless transition for which the Mott lobes closes according to
\begin{equation}
      \Delta(t) \propto A \exp\left( -\frac{W}{\sqrt{t_{\rm KT}-t}} \right) 
                          \;\; , \;\; (t_{\rm KT}-t) \ll 1 \;\; .
\label{KT}
\end{equation}
This highly nonanalytic behavior renders a direct series extrapolation
complicated and the results of such a simpleminded approach are not
conclusive.  In order to determine the phase diagram we thus performed
a Pade analysis for $\ln^2 \Delta(t) \propto (t_{\rm
KT}-t)^{-1}$. This quantity has a simple pole at the critical point
which can be captured by a rational function. 

\begin{figure}[t]
  \begin{center} 
        \epsfig{file=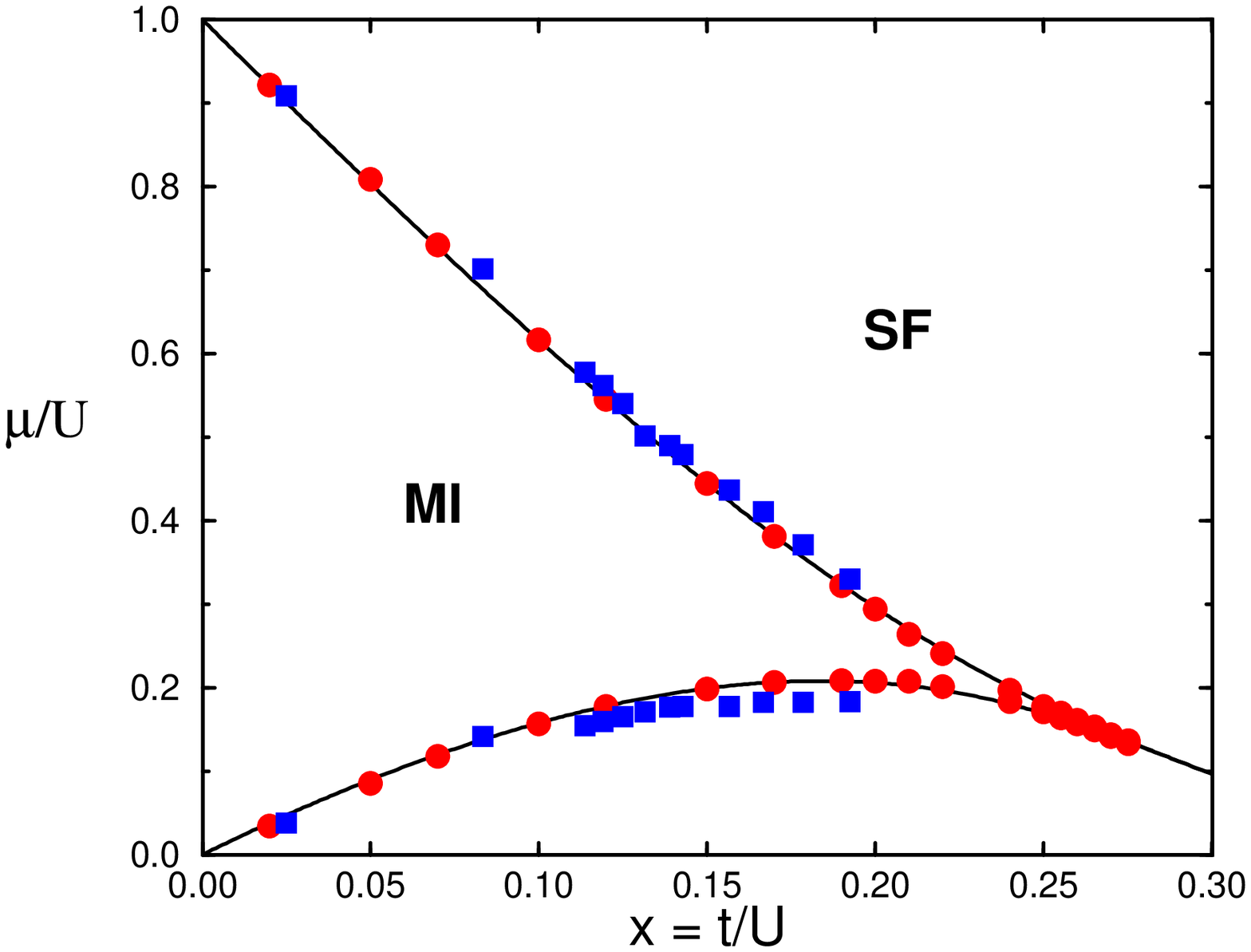,width=246pt} 
        \caption[Phase Diagram] 
                {Phase Diagram of the Bose-Hubbard chain: 
                 Numerical data from DMRG (light circles, 
                 ref.(\cite{KM98}) and QMC studies (dark squares,
                 ref.(\cite{BS92}) compared to the results of
                 a Pade analysis.}
        \label{phase-diagram:1D} 
   \end{center}
\end{figure}

The phase diagram obtained this way is shown in
Fig.\ref{phase-diagram:1D} where we compare results from the series
analysis with numerical data of QMC simulations by Batrouni and
Scalettar\cite{BS92} and a recent DMRG study\cite{K97,KM98}. The
agreement between the series and the DMRG data is excellent. Quite
remarkable is the reentrance behavior found for $t/U > 0.2$. The
series analysis thus confirms this phenomenon first observed in the DMRG
calculation. This type of behavior is a new feature not seen before in
investigations of Bose systems. 

A simple intuitive way of understanding this phenomenon is the fact
that Mott lobe is particle-hole asymmetric for the lattice problem.
Starting from strong coupling ($U \gg t$) it is clear that the effect
of the kinetic energy is to delocalize the particles. The
delocalization decreases the average number of particles per site if
the chemical potential is held fixed. On the other hand in the weak
coupling limit ($U \ll t$) the bosons condense at the lower band edge
so that for increasing bandwith ($t$) and fixed chemical potential the
average number of particles per site is decreasing. The nonmonotonic
behavior of the density can be understood simply as a result of two
limiting cases. Thus starting from the Mott phase the number of
particles per site first decreases and then increases leading to a
second Mott transition for a well defined range of the chemical
potential.
{\bf plot of $n=const.$ lines in mean field ?}

The uncertainties in the precise location of the Kosterlitz-Thouless
transition are still comparatively large. We use a Pade analysis of
$\ln^2 \Delta(t) \propto (t_{\rm KT}-t)^{-1}$.  This methods turned out 
to give excellent results.  We
estimate the point for Kosterlitz-Thouless transition to be located at
$t_{\rm KT}/U = 0.26\pm0.01$ and $\mu_{\rm KT}/U = 0.16\pm0.01$.

\section{Conclusion}

In conclusion, series expansion techniques were applied to investigate
the zero temperature properties of the Bose-Hubbard model. We
determine the complete spectrum of single-particle and single-hole
excitations in the Mott phase. The phase diagram in one and two
dimensions is obtained and the critical end points of the Mott
insulator regions are determined. In two dimensions this is so far the
only investigation of the complete phase diagram of this problem. In 
one dimensions the series shows almost perfect agreement with a 
recent DMRG study and provides a conclusive confirmation for 
reentrence behavior from the compressible to the insulating phase
near the Kosterlitz-Thouless point. 

We achnowledge useful and interesting discussions on this problem with
M.~P.~Gelfand, T.~Giamarchi, T.~K\"uhner, A.~J.~Millis, R.~Noack, 
A.~v.~Otterlo, R.~R.~P.~Singh, U.~Schollw\"ock, G.~Sch\"on, H.~Schulz, and 
S.~R.~White

\appendix

\section{Explicit Series Expansion for small clusters}

Perturbation theory for physical observables can be formulated as
a coupled cluster expansion, see e.g. \cite{GSH90}. This allows to
write the ground state energy for the Bose-Hubbard model on a particular 
lattice $\cal L$ in the following way
\begin{equation} 
      E_{\rm Mott}({\cal L})/N = E^{(0)}_{\rm Mott}/N 
                               + \sum_g L(g)\times W_E(g) \;\;\;\;. 
\label{cce-basic-eqn}
\end{equation}
Here $g$ are the connected clusters or graphs that can be embedded into the
lattice $\cal L$, the so called lattice constant $L(g)$ is the number of 
embeddings {\sl per site} of $g$ in ${\cal L}$ and $W_E(g)$, the cluster 
weight, is the contribution of $g$ to the expansion of the energy $E$. 
Each cluster represents itself a small lattice and the weights can be 
expressed by the recursion relation
\begin{equation}
      W_E(g) = E_{\rm Mott}(g) - \sum_{g'\subset g} W_E(g') \;\;\;\;,
\label{weight-eqn}
\end{equation}
where $E_{\rm Mott}(g)$ is the ground state energy of the particular
cluster $g$ under consideration and the $g'$ are all its subclusters. 
An important property of the weights is that
\begin{equation}
      W_E(g) = {\cal O}(x^n) \;\;\;\;,
\end{equation}
for a cluster with $n$ bonds ($x=t/U$). The procedure to calculate
the series for the ground state energy up to a certain order ${\cal O}(x^n)$ 
is now straightforward: first one has to find all cluster with up to
$n$ bonds and determine their lattice constants $L(g)$, in the next step 
for each cluster its ground state energy $E_{\rm Mott}(g)$ has to be 
determined perturbatively from which the weights follow by 
eqn.(\ref{weight-eqn}). Summing up all the contributions, 
eqn.(\ref{cce-basic-eqn}) gives the desired result.

As an example we briefly discuss the expansion for the Bose-Hubbard chain. 
The clusters that contribute are particular simple: finite open chains with
$n$ bonds and $n+1$ sites all having lattice constant $L(g)=1$. The first
three cluster that contribute have the following ground state energies:

\vskip 11pt
\begin{minipage}[t]{236pt}
\begin{tabular}{l c r}
 \phantom{W} $g$ & \phantom{W}\hskip 6pt\phantom{W} & $E(g)$\phantom{WW} \\
 & & \\
\begin{picture}(32,0)
      \thicklines
      \put(-5,-3){\boldmath\large$\times$}
      \put(27,-3){\boldmath\large$\times$}
      \put(0,0){\line(1,0){32}}
\end{picture} & & $-4\,x^2 + 16\,x^4 - 128\,x^6$ \\
 & & \\
\begin{picture}(64,0)
      \thicklines
      \put(-5,-3){\boldmath\large$\times$}
      \put(27,-3){\boldmath\large$\times$}
      \put(59,-3){\boldmath\large$\times$}
      \put(0,0){\line(1,0){64}}
\end{picture} & & $-8\,x^2 + 20\,x^4 + \phantom{1}\frac{40}{3}\,x^6$ \\
 & & \\
\begin{picture}(96,0)
      \thicklines
      \put(-5,-3){\large$\times$}
      \put(27,-3){\large$\times$}
      \put(59,-3){\large$\times$}
      \put(91,-3){\large$\times$}
      \put(0,0){\line(1,0){96}}
\end{picture} & & $-12\,x^2 + 24\,x^4 + \frac{392}{9}\,x^6$

\end{tabular}
\end{minipage}
\vskip 11pt

Nest step is the subcluster subtraction: the one bond cluster can be embedded
two (three) times into two (three) bond cluster and the two bond cluster 
also two times into the three bond cluster. This gives for the weights:

\vskip 11pt
\begin{minipage}[t]{236pt}
\begin{tabular}{l c r}
 \phantom{W} $g$ & \phantom{W}\hskip 6pt\phantom{W} & $W_E(g)$\phantom{WW} \\
 & & \\
\begin{picture}(32,0)
      \thicklines
      \put(0,0){\circle*{6}}
      \put(32,0){\circle*{6}}
      \put(0,0){\line(1,0){32}}
\end{picture} & & $-4\,x^2 + 16\,x^4 - \phantom{1}128\,x^6 $ \\
 & & \\
\begin{picture}(64,0)
      \thicklines
      \put(0,0){\circle*{6}}
      \put(32,0){\circle*{6}}
      \put(64,0){\circle*{6}}
      \put(0,0){\line(1,0){64}}
\end{picture} & & $ - 12\,x^4 + \phantom{1}\frac{808}{3}\,x^6$\\
 & & \\
\begin{picture}(96,0)
      \thicklines
      \put(0,0){\circle*{6}}
      \put(32,0){\circle*{6}}
      \put(64,0){\circle*{6}}
      \put(96,0){\circle*{6}}
      \put(0,0){\line(1,0){96}}
\end{picture} & & $ - \frac{1000}{9}\,x^6$

\end{tabular}
\end{minipage}
\vskip 11pt

Summing up all these different term gives the series for the ground state 
energy:
$$ E_{\rm Mott}({\cal L}=1D)/N = -4\,x^2 + 4\,x^4 + \frac{272}{9}\,x^6 $$

We now turn to the construction of the effective hamiltonian for the excited
states. As an example we will discuss the single particle states,
eqn.(\ref{eq:sp-state}). The effective hamiltonians obtained by degenerate 
perturbatition theory for the clusters with up to three bonds are given below.
Note that the energies are given with respect to the ground state energy,
because the coupled cluster theorem applies only to the difference
$H_{\rm eff} - E_0$ \cite{Gelfand:96}. 

\vskip 11pt
\begin{minipage}[t]{236pt}
\begin{picture}(32,11)
      \thicklines
      \put(-5.25,-4){\Large$\times$}
      \put(-2.5,6){1}
\end{picture}
\begin{eqnarray*}
 H(1,1) &=&  1\, \phantom{+ 0\,x + 5/2\,x^2 - 21/2\,x^3 }
\end{eqnarray*}
\begin{picture}(32,11)
      \thicklines
      \put(-5,-3){\large$\times$}
      \put(27,-3){\large$\times$}
      \put(-2.5,6){1}
      \put(29.5,6){2}
      \put(0,0){\line(1,0){32}}
\end{picture}
\begin{eqnarray*}
 H(1,1) &=& 1\, + 0\,x\, + 5/2\,x^2\, \phantom{- 21/2\,x^3} \\
 H(1,2) &=& 0\, + 2\,x\, + \phantom{5/}0\,x^2\, - \phantom{1}3/2\,x^3 \\
 H(2,1) &=& 0\, + 2\,x\, + \phantom{5/}0\,x^2\, - \phantom{1}3/2\,x^3 \\
 H(2,2) &=& 1\, + 0\,x\, + 5/2\,x^2\, \phantom{- 21/2\,x^3}
\end{eqnarray*}
\begin{picture}(64,11)
      \thicklines
      \put(-5,-3){\large$\times$}
      \put(27,-3){\large$\times$}
      \put(59,-3){\large$\times$}
      \put(-2.5,6){1}
      \put(29.5,6){2}
      \put(61.5,6){3}
      \put(0,0){\line(1,0){64}}
\end{picture}
\begin{eqnarray*}
 H(1,1) &=& 1\, + 0\,x\, + 5/2\,x^2\, \phantom{- 21/2\,x^3} \\
 H(1,2) &=& 0\, + 2\,x\, + \phantom{5/}0\,x^2\, - 21/2\,x^3 \\
 H(1,3) &=& 0\, + 0\,x\, - \phantom{5/}2\,x^2\, \phantom{- 21/2\,x^3} \\
 H(2,1) &=& 0\, + 2\,x\,   \phantom{+ 5/2\,x^2\, - 21/2\,x^3} \\
 H(2,2) &=& 1\, + 0\,x\, + \phantom{5/}5\,x^2\, \phantom{- 21/2\,x^3} \\
 H(2,3) &=& 0\, + 2\,x\,   \phantom{+ 5/2\,x^2\, - 21/2\,x^3} \\
 H(3,1) &=& 0\, + 0\,x\, - \phantom{5/}2\,x^2\, \phantom{- 21/2\,x^3} \\
 H(3,2) &=& 0\, + 2\,x\, + \phantom{5/}0\,x^2\, - 21/2\,x^3 \\
 H(3,3) &=& 1\, + 0\,x\, + 5/2\,x^2\, \phantom{- 21/2\,x^3}
\end{eqnarray*}
\end{minipage}
\vskip 11pt

\begin{minipage}[t]{236pt}
\begin{picture}(96,11)
      \thicklines
      \put(-5,-3){\large$\times$}
      \put(27,-3){\large$\times$}
      \put(59,-3){\large$\times$}
      \put(91,-3){\large$\times$}
      \put(-2.5,6){1}
      \put(29.5,6){2}
      \put(61.5,6){3}
      \put(93.5,6){4}
      \put(0,0){\line(1,0){96}}
\end{picture}
\begin{eqnarray*}
 H(1,1) &=& 1\, + 0\,x\, + 5/2\,x^2\, \phantom{- 21/2\,x^3} \\
 H(1,2) &=& 0\, + 2\,x\, + \phantom{5/}0\,x^2\, - 21/2\,x^3 \\
 H(1,3) &=& 0\, + 0\,x\, - \phantom{5/}2\,x^2\, \phantom{- 21/2\,x^3} \\
 H(1,4) &=& 0\, + 0\,x\, - \phantom{5/}0\,x^2\, + \phantom{21/}6\,x^3 \\
 H(2,1) &=& 0\, + 2\,x\,   \phantom{+ 5/2\,x^2\, - 21/2\,x^3} \\
 H(2,2) &=& 1\, + 0\,x\, + \phantom{5/}5\,x^2\, \phantom{- 21/2\,x^3} \\
 H(2,3) &=& 0\, + 2\,x\, + \phantom{5/}0\,x^2\, - \phantom{21/}9\,x^3 \\
 H(2,4) &=& 0\, + 0\,x\, - \phantom{5/}2\,x^2\, \phantom{- 21/2\,x^3} \\
 H(3,1) &=& 0\, + 0\,x\, - \phantom{5/}2\,x^2\, \phantom{- 21/2\,x^3} \\
 H(3,2) &=& 0\, + 2\,x\, + \phantom{5/}0\,x^2\, - \phantom{21/}9\,x^3 \\
 H(3,3) &=& 1\, + 0\,x\, + \phantom{5/}5\,x^2\, \phantom{- 21/2\,x^3} \\
 H(3,4) &=& 0\, + 2\,x\,   \phantom{+ 5/2\,x^2\, - 21/2\,x^3} \\
 H(4,1) &=& 0\, + 0\,x\, - \phantom{5/}0\,x^2\, + \phantom{21/}6\,x^3 \\
 H(4,2) &=& 0\, + 0\,x\, - \phantom{5/}2\,x^2\, \phantom{- 21/2\,x^3} \\
 H(4,3) &=& 0\, + 2\,x\, + \phantom{5/}0\,x^2\, - 21/2\,x^3 \\
 H(4,4) &=& 1\, + 0\,x\, + 5/2\,x^2\, \phantom{- 21/2\,x^3}
\end{eqnarray*}
\end{minipage}
\vskip 11pt

The cluster hamiltonians are not hermitian in all orders of the expansions. 
While this looks strange at first sight it is just a consequence of the 
finite cluster size which lacks the translational symmetry of the full lattice.
Therefore the sites of a finite cluster are no longer all equal and so
the sequence of intermediate virtual states that connects two different sites
$i$ and $j$ depends on the actual order, i.e. in perturbation theory
going from site $i$ to site $j$ is not equivalent to going from $j$ to $i$.
An example is the following sequence of states that contributes in
third order to $H(1,2)$ the hamiltonian of the two bond cluster:
$$ |\,2\,,\,1\,,\,1 > \; \rightarrow \; |\,2\,,\,2\,,\,0 >
\;\; \rightarrow \; |\,1\,,\,3\,,\,0 >
\;\; \rightarrow \; |\,1\,,\,2\,,\,1 > $$
In the contribution to $H(2,1)$ the same intermediate states will contribute
but in reverse order and this gives rise to the non hermitian cluster 
hamiltonian. There is no equivalent sequence to the one shown above, because
the sites 1 and 2 are not equivalent. When the effective hamiltonian for the 
full lattice will calculated later on all embeddings of the cluster into the 
lattice have to be taken into account. This restores the the translational 
symmetry and the resulting Hamiltonian will be hermitian again. 

A new complication arises when performing the subcluster subtraction, because
matrix elements have to be subtracted and it is therefore necessary 
to keep track of exactly which sites of a cluster are covered by the 
sites in a particular subcluster. The subtraction procedure is shown 
in the following graph.

\vskip 11pt
\begin{picture}(128,11)
      \thicklines
      \put(0,0){\circle*{6}}
      \put(-2.5,6){1}
      \put(0,0){\line(1,0){32}}
      \put(32,0){\circle*{6}}
      \put(29.5,6){2}
      \put(60,1){\line(1,0){8}}
      \put(60,-1){\line(1,0){8}}
      \put(91,-3){\large$\times$}
      \put(123,-3){\large$\times$}
      \put(93.5,6){1}
      \put(125.5,6){2}
      \put(96,0){\line(1,0){32}}
\end{picture}
\vskip 16pt

\begin{picture}(128,11)
      \put(61,0){\line(1,0){6}}
      \thicklines
      \put(96,0){\circle*{6}}
      \put(93.5,6){1}
      \put(89.5,-14){(1)}
      \put(128,0){\circle*{6}}
      \put(125.5,6){2}
      \put(121.5,-14){(1)}
\end{picture}
\vskip 22pt

\vskip 22pt
\begin{picture}(192,11)
      \thicklines
      \put(0,0){\circle*{6}}
      \put(-2.5,6){1}
      \put(0,0){\line(1,0){32}}
      \put(32,0){\circle*{6}}
      \put(29.5,6){2}
      \put(32,0){\line(1,0){32}}
      \put(64,0){\circle*{6}}
      \put(61.5,6){3}
      \put(92,1){\line(1,0){8}}
      \put(92,-1){\line(1,0){8}}
      \put(123,-3){\large$\times$}
      \put(155,-3){\large$\times$}
      \put(187,-3){\large$\times$}
      \put(125.5,6){1}
      \put(157.5,6){2}
      \put(189.5,6){3}
      \put(128,0){\line(1,0){64}}
\end{picture}
\vskip 16pt

\begin{picture}(192,11)
      \put(93,0){\line(1,0){6}}
      \thicklines
      \put(128,0){\circle*{6}}
      \put(125.5,6){1}
      \put(121.5,-14){(1)}
      \put(160,0){\circle*{6}}
      \put(157.5,6){2}
      \put(153.5,-14){(1)}
      \put(192,0){\circle*{6}}
      \put(189.5,6){3}
      \put(185.5,-14){(1)}
\end{picture}
\vskip 22pt

\vskip 11pt
\begin{picture}(192,11)
      \put(93,0){\line(1,0){6}}
      \thicklines
      \put(128,0){\circle*{6}}
      \put(125.5,6){1}
      \put(121.5,-14){(1)}
      \put(128,0){\line(1,0){32}}
      \put(160,0){\circle*{6}}
      \put(157.5,6){2}
      \put(153.5,-14){(2)}
      \put(192,0){\circle{6}}
      \put(189.5,6){3}
\end{picture}
\vskip 22pt

\vskip 11pt
\begin{picture}(192,11)
      \put(93,0){\line(1,0){6}}
      \thicklines
      \put(128,0){\circle{6}}
      \put(125.5,6){1}
      \put(160,0){\circle*{6}}
      \put(157.5,6){2}
      \put(153.5,-14){(1)}
      \put(160,0){\line(1,0){32}}
      \put(192,0){\circle*{6}}
      \put(189.5,6){3}
      \put(185.5,-14){(2)}
\end{picture}
\vskip 22pt

The subtraction procedure gives the cluster weights of the hamiltonian and
as it is the case for the ground state energy the weights for a cluster
with $n$-bonds are at least of order ${\cal O}(x^n)$.

\vskip 11pt
\begin{minipage}[t]{236pt}
\begin{picture}(32,11)
      \thicklines
      \put(0,0){\circle*{6}}
      \put(-2.5,6){1}
\end{picture}
\begin{eqnarray*}
 W_H(1,1) &=&  1\, \phantom{+ 0\,x + 5/2\,x^2 - 21/2\,x^3}
\end{eqnarray*}
\begin{picture}(32,11)
      \thicklines
      \put(0,0){\circle*{6}}
      \put(-2.5,6){1}
      \put(0,0){\line(1,0){32}}
      \put(32,0){\circle*{6}}
      \put(29.5,6){2}
\end{picture}
\begin{eqnarray*}
 W_H(1,1) &=& 0\, + 0\,x\, + 5/2\,x^2\, \phantom{- 21/2\,x^3} \\
 W_H(1,2) &=& 0\, + 2\,x\, + \phantom{5/}0\,x^2\, - \phantom{1}3/2\,x^3 \\
 W_H(2,1) &=& 0\, + 2\,x\, + \phantom{5/}0\,x^2\, - \phantom{1}3/2\,x^3 \\
 W_H(2,2) &=& 0\, + 0\,x\, + 5/2\,x^2\, \phantom{- 21/2\,x^3}
\end{eqnarray*}
\begin{picture}(64,11)
      \thicklines
      \put(0,0){\circle*{6}}
      \put(-2.5,6){1}
      \put(0,0){\line(1,0){32}}
      \put(32,0){\circle*{6}}
      \put(29.5,6){2}
      \put(32,0){\line(1,0){32}}
      \put(64,0){\circle*{6}}
      \put(61.5,6){3}
\end{picture}
\begin{eqnarray*}
 W_H(1,2) &=& 0\, + 0\,x\, + \phantom{5/}0\,x^2\, - \phantom{21/}9\,x^3 \\
 W_H(1,3) &=& 0\, + 0\,x\, - \phantom{5/}2\,x^2\, + \phantom{21/}0\,x^3 \\
 W_H(2,1) &=& 0\, + 0\,x\, + \phantom{5/}0\,x^2\, + \phantom{1}3/2\,x^3 \\
 W_H(2,3) &=& 0\, + 0\,x\, + \phantom{5/}0\,x^2\, + \phantom{1}3/2\,x^3 \\
 W_H(3,1) &=& 0\, + 0\,x\, - \phantom{5/}2\,x^2\, + \phantom{21/}0\,x^3 \\
 W_H(3,2) &=& 0\, + 0\,x\, + \phantom{5/}0\,x^2\, - \phantom{21/}9\,x^3 \\
\end{eqnarray*}
\begin{picture}(96,11)
      \thicklines
      \put(0,0){\circle*{6}}
      \put(-2.5,6){1}
      \put(0,0){\line(1,0){32}}
      \put(32,0){\circle*{6}}
      \put(29.5,6){2}
      \put(32,0){\line(1,0){32}}
      \put(64,0){\circle*{6}}
      \put(61.5,6){3}
      \put(64,0){\line(1,0){32}}
      \put(96,0){\circle*{6}}
      \put(93.5,6){4}
\end{picture}
\begin{eqnarray*}
 W_H(1,4) &=& 0\, + 0\,x\, + \phantom{5/}0\,x^2\, + \phantom{21/}6\,x^3 \\
 W_H(4,1) &=& 0\, + 0\,x\, + \phantom{5/}0\,x^2\, + \phantom{21/}6\,x^3 
\end{eqnarray*}
\end{minipage}
\vskip 11pt

The last steps requires to find all possible embeddings of the cluster 
in the chain. Due to the translational invariance of the problem, it is
sufficient to search only those possibilities where that connect the origin 
to other sites. For the first clusters this is demonstrated in the following:

\vskip 11pt
\begin{picture}(96,11)
      \thicklines
      \put(0,0){\circle{6}}
      \put(-4,6){-3}
      \put(32,0){\circle{6}}
      \put(28,6){-2}
      \put(64,0){\circle{6}}
      \put(60,6){-1}
      \put(96,0){\circle*{6}}
      \put(93.5,6){0}
      \put(89.5,-14){(1)}
      \put(128,0){\circle{6}}
      \put(125.5,6){1}
      \put(160,0){\circle{6}}
      \put(157.5,6){2}
      \put(192,0){\circle{6}}
      \put(189.5,6){3}
\end{picture}
\vskip 22pt

\vskip 11pt
\begin{picture}(96,11)
      \thicklines
      \put(0,0){\circle{6}}
      \put(-4,6){-3}
      \put(32,0){\circle{6}}
      \put(28,6){-2}
      \put(64,0){\circle{6}}
      \put(60,6){-1}
      \put(96,0){\circle*{6}}
      \put(96,0){\line(1,0){32}}
      \put(93.5,6){0}
      \put(89.5,-14){(1)}
      \put(128,0){\circle*{6}}
      \put(125.5,6){1}
      \put(121.5,-14){(2)}
      \put(160,0){\circle{6}}
      \put(157.5,6){2}
      \put(192,0){\circle{6}}
      \put(189.5,6){3}
\end{picture}
\vskip 22pt

\vskip 11pt
\begin{picture}(96,11)
      \thicklines
      \put(0,0){\circle{6}}
      \put(-4,6){-3}
      \put(32,0){\circle{6}}
      \put(28,6){-2}
      \put(64,0){\circle*{6}}
      \put(64,0){\line(1,0){32}}
      \put(60,6){-1}
      \put(57.5,-14){(1)}
      \put(96,0){\circle*{6}}
      \put(93.5,6){0}
      \put(89.5,-14){(2)}
      \put(128,0){\circle{6}}
      \put(125.5,6){1}
      \put(160,0){\circle{6}}
      \put(157.5,6){2}
      \put(192,0){\circle{6}}
      \put(189.5,6){3}
\end{picture}
\vskip 22pt

\vskip 11pt
\begin{picture}(96,11)
      \thicklines
      \put(0,0){\circle{6}}
      \put(-4,6){-3}
      \put(32,0){\circle{6}}
      \put(28,6){-2}
      \put(64,0){\circle{6}}
      \put(60,6){-1}
      \put(96,0){\circle*{6}}
      \put(96,0){\line(1,0){32}}
      \put(93.5,6){0}
      \put(89.5,-14){(1)}
      \put(128,0){\circle*{6}}
      \put(128,0){\line(1,0){32}}
      \put(125.5,6){1}
      \put(121.5,-14){(2)}
      \put(160,0){\circle*{6}}
      \put(157.5,6){2}
      \put(153.5,-14){(3)}
      \put(192,0){\circle{6}}
      \put(189.5,6){3}
\end{picture}
\vskip 22pt

\vskip 11pt
\begin{picture}(96,11)
      \thicklines
      \put(0,0){\circle{6}}
      \put(-4,6){-3}
      \put(32,0){\circle{6}}
      \put(28,6){-2}
      \put(64,0){\circle*{6}}
      \put(64,0){\line(1,0){32}}
      \put(60,6){-1}
      \put(57.5,-14){(1)}
      \put(96,0){\circle*{6}}
      \put(96,0){\line(1,0){32}}
      \put(93.5,6){0}
      \put(89.5,-14){(2)}
      \put(128,0){\circle*{6}}
      \put(125.5,6){1}
      \put(121.5,-14){(3)}
      \put(160,0){\circle{6}}
      \put(157.5,6){2}
      \put(192,0){\circle{6}}
      \put(189.5,6){3}
\end{picture}
\vskip 22pt

\vskip 11pt
\begin{picture}(96,11)
      \thicklines
      \put(0,0){\circle{6}}
      \put(-4,6){-3}
      \put(32,0){\circle*{6}}
      \put(32,0){\line(1,0){32}}
      \put(28,6){-2}
      \put(25.5,-14){(1)}
      \put(64,0){\circle*{6}}
      \put(64,0){\line(1,0){32}}
      \put(60,6){-1}
      \put(57.5,-14){(2)}
      \put(96,0){\circle*{6}}
      \put(93.5,6){0}
      \put(89.5,-14){(3)}
      \put(128,0){\circle{6}}
      \put(125.5,6){1}
      \put(160,0){\circle{6}}
      \put(157.5,6){2}
      \put(192,0){\circle{6}}
      \put(189.5,6){3}
\end{picture}
\vskip 22pt

Up to third order in the expansion parameter the effective hamiltonian 
for the single particle states in the Bose-Hubbard chain reads:
\begin{eqnarray*}
H_{\rm eff}(-3) &=& 0\, + 0\,x\, + 0\,x^2\, + 6\,x^3 \\
H_{\rm eff}(-2) &=& 0\, + 0\,x\, - 2\,x^2\, + 0\,x^3 \\
H_{\rm eff}(-1) &=& 0\, + 2\,x\, + 0\,x^2\, - 9\,x^3 \\
H_{\rm eff}(0)  &=& 1\, + 0\,x\, + 5\,x^2\, + 6\,x^3 \\
H_{\rm eff}(1)  &=& 0\, + 2\,x\, + 0\,x^2\, - 9\,x^3 \\
H_{\rm eff}(2)  &=& 0\, + 0\,x\, - 2\,x^2\, + 0\,x^3 \\
H_{\rm eff}(3)  &=& 0\, + 0\,x\, + 0\,x^2\, + 6\,x^3
\end{eqnarray*}
Note that the effective hamiltonian is hermitian. 

\section{Hash function for Lattice Bosons}

Perturbation theory requires to work in the full hilbert space
of the finite clusters. It is therefore necessary to have an
efficient method to represent the basis and the hamiltonian. 
In  particular there is a need for a Hash function that labels
each basis state in a unique way. The hamiltonian (\ref{bh-model}) 
conserves the total particle number $N$. For a cluster with $L$ sites
and fixed $N$ the hilbert space has a total of
\begin{equation}
D(N,L) = \frac{(N+L-1)!}{N! \, (L-1)!}
\end{equation}
states. The natural basis is given by the set of occupation numbers $\{n_i\}$ 
of all lattice sites $i$, 
$ | \, n_1 \,,\, n_2 \,,\, n_3 \,,\, ... \,,\, n_{\rm L} \, >$ .
We now describe how to construct the full basis. 
The first state is the one where all particles are sitting on site $1$. 
Then follow those states were the first two sites are occupied. 
In the next stepthe last occupied site is $i=3$. From those possibilities 
we first get all those with just one boson on site $3$. There are then three
bosons left to be distributed on $i=1$ and $i=2$ and we repeat the procedure, 
first all three on site 1 and so on.

Define
\begin{equation}
\theta_i := \sum_{j=1}^i n_i \;\;\;\;, \;\;\;\; \theta_0 := 0
\end{equation}
Which is related to the occupation number by
\begin{equation}
n_i = \theta_i -  \theta_{i-1}
\end{equation}
Hash function
\begin{equation}
Hf(\{n_i\})  = \sum_{j=1}^L 
               \,\left[\phantom{^i}D(\theta_i,i) - D(\theta_{i-1},i) \,\right]
\end{equation}

\begin{table}
\label{series:sqr}
\caption{Series for the single particle gap $\Delta$ annd the equal time
structure factor $S_{{\bf q}=0}$ of the Bose-Hubbard model on 
the square lattice.}
\begin{tabular}{ c | r }
\multicolumn{2}{c}{gap series} \\
\hline
  n  &    $a_n$  \\
\hline 
           0  &  1  \\
           1  &  -12  \\
           2  &  -22  \\
           3  &  -264  \\
           4  &  -15659\,/\,10  \\
           5  &  -656984\,/\,25  \\
           6  &  -513092341\,/\,2250  \\
           7  &  -13396365654\,/\,3375  \\
           8  &  -2194497431888101\,/\,56700000  \\
           9  &  -523244582353596437\,/\,744187500  \\
          10  &  -4749112579154967367231\,/\,625117500000  \\
          11  &  -6676218845916748474723399\,/\,49228003125000  \\
          12  &  -5669114326328304841982042447\,/\,3508972062750000  \\
          13  &  -3473317126780784521271398100\,/\,126449317253259  \\
\hline
\hline
\multicolumn{2}{c}{structure factor series} \\
\hline
  n  &    $a_n$  \\
\hline 
           0  &  3  \\
           1  &  32  \\
           2  &  432  \\
           3  &  6656  \\
           4  &  99632  \\
           5  &  14154496\,/\,9  \\
           6  &  663550400\,/\,27  \\
           7  &  31905307840\,/\,81  \\
           8  &  7618958766796\,/\,1215  \\
           9  &  12934227681606432\,/\,127575  \\
          10  &  14570617373829713351\,/\,8930250  \\
          11  &  19905912307529372064253\,/\,750141000  \\
          12  &  17231660529006598072716626038\,/\,40068302174901  \\
          13  &  47514032474492554578981737799.4028\,/\,6764778289269  \\
\end{tabular}
\end{table}

\begin{table}
\label{series:tri}
\caption{Series for the single particle gap $\Delta$ annd the equal time
structure factor $S_{{\bf q}=0}$ of the Bose-Hubbard model on 
the triangular lattice.}
\begin{tabular}{ c | r }
\multicolumn{2}{c}{gap series} \\
\hline
  n  &    $a_n$  \\
\hline 
           0  &  1  \\
           1  &  -18  \\
           2  &  -81  \\
           3  &  -819  \\
           4  &  -54891\,/\,4  \\
           5  &  -11459377\,/\,50  \\
           6  &  -6764830501\,/\,1500  \\
           7  &  -3957593443549\,/\,45000  \\
           8  &  -23724030434424597\,/\,12600000  \\
           9  &  -79054659543137812691\,/\,1984500000  \\
          10  &  -49507676563116513700717\,/\,55566000000  \\
          11  &  -19752788544107.0312336222045518956  \\
          12  &  -454652221307484.008486685913608139  \\
          13  &  -10398395856680122.5582305335816799  \\
\hline
\hline
\multicolumn{2}{c}{structure factor series} \\
\hline
  n  &    $a_n$  \\
\hline 
           0  &  3  \\
           1  &  48  \\
           2  &  1080  \\
           3  &  25344  \\
           4  &  613016  \\
           5  &  15108128  \\
           6  &  3391856144\,/\,9  \\
           7  &  28444402112\,/\,3  \\
           8  &  97228564590772\,/\,405  \\
           9  &  86595555599744452\,/\,14175  \\
          10  &  2787620342817465447617\,/\,17860500  \\
          11  &  82527451969123616224435919\,/\,20628877500  \\
          12  &  102821325219551430.224270929006395  \\
          13  &  2648736908451106507.61278132531748   \\
\end{tabular}
\end{table}

\begin{table}
\label{series:1D}
\caption{Series for the single particle gap $\Delta$ annd the equal time
structure factor $S_{{\bf q}=0}$ of the Bose-Hubbard chain.}
\begin{tabular}{ c | r }
\multicolumn{2}{c}{gap series} \\
\hline
  n  &    $a_n$  \\
\hline 
           0  &  1  \\
           1  &  -6  \\
           2  &   5 \\
           3  &   6 \\
           4  &  287\,/\,20  \\
           5  &  17463\,/\,150  \\
           6  &  -1806729\,/\,3000  \\
           7  &  73674531\,/\,22500  \\
           8  &  -297690613629\,/\,16200000  \\
           9  &  14666046468323\,/\,121500000  \\
          10  &  -6295148943458549\,/\,7290000000  \\
          11  &  114441271150219589\,/\,18225000000  \\
          12  &  -422231271662550684871\,/\,9185400000000  \\
          13  &  1473292023890353319230511\,/\,4340101500000000  \\
\hline
\hline
\multicolumn{2}{c}{structure factor series} \\
\hline
  n  &    $a_n$  \\
\hline 
           0  &  3  \\
           1  &  16  \\
           2  &  72  \\
           3  &  320  \\
           4  &  4120\,/\,3  \\
           5  &  48544\,/\,9  \\
           6  &  596992\,/\,27  \\
           7  &  2396512\,/\,27  \\
           8  &  27653840\,/\,81  \\
           9  &  1329678  \\
          10  &  22598877209\,/\,4375  \\
          11  &  1275342277201\,/\,65610  \\
          12  &  4055776421430107\,/\,55112400  \\
          13  &  203597243119484303\,/\,723350250   \\
\end{tabular}
\end{table}

\begin{table}
\label{pade:sqr}
\caption{D-log Pade approximants for the critical point $t_c$ and the 
critical exponents $\nu$ and $\gamma_s$ of the Bose-Hubbard model on 
the square lattice.}
\begin{tabular}{ c | c c | c c }
        & \multicolumn{2}{c|}{gap}   &\multicolumn{2}{c}{structure factor} \\
\hline
  [L/M]  &    $t_c$     &    $\nu$     &   $t_c$      &  $\gamma_s$  \\
\hline 
  $\rm [4/4]$  &  0.05980859  &  0.69137519  &  0.05976303  &  0.65480784  \\
  $\rm [4/5]$  &  0.05978709  &  0.68924487  &  0.05976978  &  0.65534864  \\
  $\rm [5/4]$  &  0.05980549  &  0.69112808  &  0.05971761  &  0.65393342  \\
  $\rm [4/6]$  &  0.05980887  &  0.69140624  &  0.05976747  &  0.65515773  \\
  $\rm [5/5]$  &  0.05980805  &  0.69133305  &  0.05976666  &  0.65508534  \\
  $\rm [6/4]$  &  0.05980880  &  0.69140183  &  0.05976814  &  0.65522423  \\
  $\rm [4/7]$  &  0.05980331  &  0.69097429  &  0.05975849  &  0.65433135  \\
  $\rm [5/6]$  &  0.05984310  &  0.69384313  &  0.05978395  &  0.65596948  \\
  $\rm [6/5]$  &  0.05980488  &  0.69108539  &  0.05976313  &  0.65477892  \\
  $\rm [7/4]$  &  0.05980859  &  0.69138480  &  0.05976778  &  0.65519366  \\
  $\rm [4/8]$  &  0.05978512  &  0.68975552  &  0.05974697  &  0.65294432  \\
  $\rm [5/7]$  &  0.05987138  &  0.69556354  &  0.05973294  &  0.65050768  \\
  $\rm [6/6]$  &  0.05987084  &  0.69552706  &  0.05972443  &  0.64864666  \\
  $\rm [7/5]$  &  0.05978277  &  0.68960137  &  0.05975399  &  0.65386465  \\
  $\rm [8/4]$  &  0.06016134  &  0.72645361  &  0.05974142  &  0.65212846
\end{tabular}  
\end{table}

\begin{table}
\label{pade:tri}
\caption{D-log Pade approximants for the critical point $t_c$ and the 
critical exponents $\nu$ and $\gamma_s$ of the Bose-Hubbard model on 
the triangular lattice.}
\begin{tabular}{ c | c c | c c }
        & \multicolumn{2}{c|}{gap}   &\multicolumn{2}{c}{structure factor} \\
\hline
  [L/M]        &    $t_c$     &    $\nu$     &    $t_c$     &  $\gamma_s$  \\
\hline 
  $\rm [4/4]$  &  0.03781415  &  0.68753874  &  0.03768377  &  0.60598248  \\
  $\rm [4/5]$  &  0.03782464  &  0.68812741  &  0.03778221  &  0.65202471  \\
  $\rm [5/4]$  &  0.03777127  &  0.67855607  &  0.03778121  &  0.65179657  \\
  $\rm [4/6]$  &  0.03780279  &  0.68619855  &  0.03777688  &  0.65066891  \\
  $\rm [5/5]$  &  0.03780194  &  0.68605142  &  0.03777941  &  0.65137574  \\
  $\rm [6/4]$  &  0.03780562  &  0.68666551  &  0.03777913  &  0.65130264  \\
  $\rm [4/7]$  &  0.03779156  &  0.68397787  &  0.03777917  &  0.65130976  \\
  $\rm [5/6]$  &  0.03776841  &  0.67611321  &  0.03777930  &  0.65134670  \\
  $\rm [6/5]$  &  0.03779251  &  0.68419813  &  0.03777930  &  0.65134598  \\
  $\rm [7/4]$  &  0.03778693  &  0.68276629  &  0.03777931  &  0.65134883  \\
  $\rm [4/8]$  &  0.03778359  &  0.68180328  &  0.03777864  &  0.65116474  \\
  $\rm [5/7]$  &  0.03778045  &  0.68074745  &  0.03777854  &  0.65113450  \\
  $\rm [6/6]$  &  0.03778034  &  0.68070944  &  0.03777943  &  0.65137873  \\
  $\rm [7/5]$  &  0.03778358  &  0.68180026  &  0.03777915  &  0.65130775  \\
  $\rm [8/4]$  &  0.03778162  &  0.68114199  &  0.03777683  &  0.65058408 
\end{tabular}  
\end{table}


\begin{references}
  \bibliographystyle{prsty}
  


\bibitem{SBZ91} R.~T.~Scalettar, G.~G.~Batrouni and G.~T.~Zimany,
    Phys. Rev. Lett {\bf 66}, 3144 (1991)

\bibitem{BS92} G.~G.~Batrouni and R.~T.~Scalettar, 
    Phys. Rev. B {\bf 46}, 9051 (1992)

\bibitem{NSFG94} P. Niyaz, R.~T. Scalettar, C.~Y. Fong and G.~G.~Batrouni, 
    Phys. Rev. B {\bf 50}, 362 (1994)

\bibitem{KT91} W. Krauth and N.~Trivedi, 
    Europhys. Lett {\bf 14}, 627 (1991)

\bibitem{KTC91} W. Krauth, N.~Trivedi and D. Ceperley,
   Phys. Rev. Lett {\bf 67}, 2307 (1991)

\bibitem{OW94} A.~van~Otterlo and K.-H. Wagenblast,
   Phys. Rev. Lett {\bf 72}, 3598 (1994)

\bibitem{BSZK95} G.~G.~Batrouni, R.~T.~Scalettar, G.~T.~Zimany and A.~P.~Kampf,
   Phys. Rev. Lett {\bf 72}, 3598 (1994)

\bibitem{GSH90} M.~P.~Gelfand, R.~R.~P.~Singh, and D.~A.~Huse, 
                J.~Stat.~Phys. {\bf 59}, 1093 (1990)

\bibitem{Gelfand:96} M.~P.~Gelfand, Sol.~Stat.~Com. {\bf 98}, 11 (1996)

\bibitem{Fisher:1989} M.~P.~A. Fisher, P.~B. Weichman, G. Grinstein,
  and D.~S. Fisher, Phys. Rev. B {\bf 40}, 546 (1989).

\bibitem{K97} T.~K\"uhner. diploma thesis, Bonn University (1997)

\bibitem{KM98} T.~K\"uhner, and H.~Monien,
               preprint, cond-mat/9712307

\bibitem{GZ77} J.~C.~Le ~Guillou, and J.~Zinn-Justin, 
                Phys.~Rev.~Lett. {\bf 39}, 95 (1977)


\bibitem{Orr.Haviland.Jaeger} B.~G. Orr, H.~M. Jaeger, A.~M. Goldman,
  and C.~G. Kuper, Phys. Rev. Lett. {\bf 56}, 378 (1986); D.~B.
  Haviland, Y. Liu, and A.~M. Goldman, Phys. Rev. Lett. {\bf {\bf
      62}}, 2180 (1989); H.~M. Jaeger, D.~B. Haviland, B.~G. Orr, and
  A.~M. Goldman, Phys. Rev. B {\bf 40}, 182 (1989).
    
  
\bibitem{Oudenaarden:1996} A. Oudenaarden and J.~E. Mooij, Phys. Rev.
  Lett. {\bf 76}, 4947 (1996).
  
\bibitem{Normalphase} R. Baltin and K.-H. Wagenblast, Europhys. Lett.
  {\bf 39}, 7 (1997); L.~I. Glazman and A.~I. Larkin, Phys. Rev.
  Lett. {\bf 79}, 3736 (1997).
  
                      
  
\bibitem{White:1992} S.~R. White, Phys. Rev. Lett. {\bf 69}, 2863 (1992).

  
\bibitem{Comment2} Pai et al. \cite{Pai:1996} have chosen a cut-off of
  four particle per site for the same phase. Kashurnikov and Svistunov
  \cite{Kashurnikov:1996} found $n=3$ to be very close to the full
  model.

  
\bibitem{BKT} V.~L. Berezinskii, Zh. Eksp. Teor. Fiz. {\bf 61},
  1144 (1971) [JETP {\bf 34}, 610 (1972)]; J.~M. Kosterlitz and
  D.~J. Thouless, Journal of Physics C {\bf 6}, 1181 (1973);
  J.~M. Kosterlitz, Journal of Physics C {\bf 7}, 1046 (1974).

  
\bibitem{Elesin:1994} V.~F. Elesin, V.~A. Kashurnikov, and L.~A.
  Openov, Pis'ma Zh. Eksp. Teor. Fiz.  {\bf 60}, 174 (1994)
  [JETP Lett. {\bf 60}, 177 (1994)].
  
\bibitem{Kashurnikov:1996.2} V.~A. Kashurnikov, A.~V. Krasavin, and
  B.~V. Svistunov, Pis'ma Zh. Eksp. Theo.  Fiz. {\bf 64}, 92
  (1996) [JETP Lett. {\bf 64}, 99 (1996)].
  
\bibitem{Pai:1996} R.~V. Pai, R. Pandit, H.~R. Krishnamurthy, and S.
  Ramasesha, Phys. Rev. Lett.  {\bf 76}, 2937 (1996).
  
\bibitem{Haldane:1981} F.~D.~M. Haldane, Physics Review Letters {\bf 47}, 
  1840 (1981); F.~D.~M. Haldane, J.Phys.C {\bf 14}, 2585 (1981).
  
\bibitem{Giamarchi:1992.2} T. Giamarchi and A.~J. Millis, Phys. Rev. B
  {\bf 46}, 9325 (1992).

\bibitem{Krauth:1991} W. Krauth, Phys.Rev.B {\bf 44}, 9772
  (1991).
  
\bibitem{Kashurnikov:1996} V.~A. Kashurnikov and B.~V. Svistunov,
  Phys. Rev. B {\bf 53}, 11776 (1996).
  
\bibitem{Batrouni:1992} G.~G. Batrouni and R.~T. Scalettar, Phys. Rev.
  B {\bf 46}, 9051 (1992).
  
\bibitem{Monien:1996} J.~K. Freericks and H. Monien, Phys. Rev. B {\bf 53}, 
  2691 (1996).
              
\bibitem{MeiselBatrouni} M.~W. Meisel, Physica B {\bf 178}, 121 (1992); 
  G.~G. Batrouni, R.~T. Scalettar, G.~T.  Zimanyi, and A.~P.
  Kampf, Phys. Rev.  Lett. {\bf 74}, 2527 (1995).
  

\end{references}
\end{document}